\documentclass[11pt]{article} 

  \usepackage{latexsym}
  \usepackage{amssymb}
  \usepackage[dvips]{graphicx}
\usepackage{rotating}

  \def\AFOUR{
  \setlength{\textheight}{9.0in}
  \setlength{\textwidth}{5.75in}

  \setlength{\topmargin}{-0.375in}
  \hoffset=-0.5in 
  \renewcommand{\baselinestretch}{1.17}
  }

  \AFOUR 

\renewcommand{\theequation}{\thesection.\arabic{equation}}

\newtheorem{theorem}{Theorem}[section]
\newtheorem{lemma}{Lemma}[section]

\begin{document}

 \begin{titlepage}
         \begin{center}
 \vskip .5in
 \begin{flushright}  ICMPA-MPA/2008/16 \\
LPT-Orsay-08-54
 \end{flushright}

 \begin{center}
 \vskip .5in

 {\LARGE\bf  Vanishing $\beta$-function for
Grosse-Wulkenhaar model in a magnetic field\footnote{Work supported
by the ANR Program ``GENOPHY" and by the  Daniel Iagolnitzer Foundation, France.} }
 \end{center}

  \begin{center}
 {\bf Joseph Ben Geloun$^{a,b,\dag}$, Razvan Gurau$^{c,\ddag}$ \\
and Vincent Rivasseau$^{c,\star}$ }\\

$^{a}$International Chair of Mathematical Physics and Applications\\ 
ICMPA-UNESCO Chair, Universit\'e d'Abomey-Calavi,
Cotonou, Rep. of Benin\\
$^{b}$Facult\'e des Sciences et Techniques \\
Universit\'e Cheikh Anta Diop, Dakar, S\'en\'egal\\
$^{c}$Laboratoire de Physique Th\'eorique, UMR CNRS 8627\\
Universit\'e Paris-Sud XI, 91405 Orsay, France\\
Email: $^\dag$joseph.bengeloun@cipma.uac.bj, \ 
$^\ddag$Razvan.Gurau@th.u-psud.fr, \
$^\star$Vincent.Rivasseau@th.u-psud.fr 
 \vspace{0.5cm}

    \end{center} 

\vspace{10pt}
 \today
 \begin{abstract}
We prove that the $\beta$-function of the Grosse-Wulkenhaar model including 
a magnetic field vanishes at all order of perturbations.
We compute the renormalization group flow of the relevant dynamic
parameters and find a non-Gaussian infrared fixed point.
Some consequences of these results are discussed.
 \end{abstract}

\noindent
\end{center}
Pacs numbers: 11.10.Nx, 11.10.Gh, 02.40.Gh.
 \end{titlepage}
 \vspace{500pt}

\section{Introduction}
\label{Sect1}

The Grosse-Wulkenhaar model (GWm) has been shown renormalizable
at all orders of perturbations \cite{GW}. This noncommutative (NC) $\phi^{\star 4 }_4$ scalar
theory has been extensively studied in the very recent years \cite{Riv}-\cite{diser2}. 
Its most remarkable property, which holds both for the real and complex version,
is asymptotic safeness.  The $\beta$-function vanishes at all order of perturbations.
This result has been proved in \cite{diser2} by a careful combination of 
Dyson-Schwinger equations (DSe) and Ward identities (WI's).
Aside from ${\cal N}=4$ supersymmetric non Abelian gauge theory, 
it is the only yet known four dimensional quantum field theory with this property.

It is natural to wonder whether there exist other NC renormalizable models 
with this striking property. The Gross-Neveu model \cite{Vign} and the
$O(N)$ and $U(N)$ invariant GWm's \cite{BGR},  when preserving Langmann-Szabo (LS) duality \cite{LS},
have been also proved renormalizable. However the Gross-Neveu model is not asymptotically safe \cite{VW},
and in the class of NC color models
considered  in \cite{BGR}, only the GWm itself (real or complex) is asymptotically safe\footnote{
Nevertheless, a large class of such models is UV asymptotic free hence
susceptible of a full constructive analysis \cite{lRiv}.}.

In this paper, we consider the complex GWm with added magnetic field, namely the
Langmann-Szabo-Zarembo (LSZ) model \cite{LSZ}.
Langmann and co-workers proposed
that small perturbations of this theory could produce solvable models with renormalizable interactions.
Here, we study the RG flow of the coupling constant, still with $\Omega=1$ and 
a magnetic field satisfying $|B| < 1$.  We prove that this model is still asymptotically safe at all orders,
and we calculate the RG flows of the two corresponding wave function constants $q$ and  $p$.
Note that beyond its possible relevance for high energy physics,
this LSZ model is a toy version of the quantum Hall effect, which can be considered
a  $2+1$ NC quantum field theory  with non relativistic propagator including 
Matsubara frequencies and Fermi surface \cite{Riv2}.

The paper is organized as follows. The next section introduces the GWm in a magnetic
field. The main theorem on the vanishing $\beta$-function and 
its proof are developed in Section 3. 
Section 4 provides derivations of RG flows
of the new parameters $q$ and $p$.
Further issues and conclusions are drawn in Section 5, while an appendix summarizes 
some calculations.

\section{Notations and considerations}

The following action describes the complex  NC $\phi^{\star 4}_4$ 
LSZ model in the Moyal-Euclidean space \cite{LSZ}
\begin{equation}
S= \int d^4x\,\left\{
\partial_{\mu}\bar \phi  \partial^{\mu}\phi +\mu^2 \bar \phi \phi
+ \Omega \bar \phi  \tilde{x}_{\mu} \tilde{x}^{\mu}\phi     +2i\,B\bar\phi (\tilde{x}_{\mu} \partial^{\mu}) \phi
 + \frac{\lambda}{2} \bar\phi \star \phi\star\bar\phi\star\phi \right\},
\end{equation}
where $B$ is the magnetic field, $\tilde{x}_\nu = 2(\theta_{\nu\mu}^{-1})x^{\mu}$ and $\theta_{\nu\mu}^{-1}$ is the inverse
of the antisymmetric matrix associated with the Moyal $\star$-product.
The mass parameter is  $\mu$ and $\Omega-B^2$ is an harmonic potential. The complex GWm
is recovered at $B=0$.

We use the matrix basis and the notations of Refs.\cite{diser,diser2}, setting
$\Omega=1$\footnote{The corresponding model is an independent non 
identically distributed matrix model.}. As argued in \cite{diser,diser2},  the 
flow of $\Omega$ goes rapidly to $1$ in the UV limit.
After rescaling the field and the coupling constant by two constants,
the generating functional becomes
\begin{eqnarray}
&&Z(\eta,\bar{\eta})=\int d\phi d\bar{\phi}~e^{-S(\bar{\phi},\phi)+F(\bar{\eta},\eta,;\bar{\phi},\phi)}, \\
&&S(\bar{\phi},\phi)=\bar{\phi}\,X_R \,\phi+\phi\, X_L \,\bar\phi+A\bar{\phi}\phi+
\frac{\lambda}{2}\phi\bar{\phi}\phi\bar{\phi},\quad
F(\bar{\eta},\eta;\bar{\phi},\phi)=  \bar{\phi}\eta+\bar{\eta}\phi, \label{action}\\
&&\phi=(\phi_{mn}),\;\; X_L= q\, m\, \delta_{mn}, \;\;\;  X_R= p\, m\, \delta_{mn};\;\;\;
q= 1+B, \;\;\;p= 1-B,
\end{eqnarray}
where traces are implicit, $S$ is the action and $F$ represents external sources.
The quadratic part of the action is now expressed in term of the left and right
matrix operators $X_L$ and $X_R$ with unequal weights $q$ and $p$, respectively,
and the new mass parameter is $A:=2+\mu^2\theta/4$ \cite{theraz}.

The theory is stable for $|B|<\Omega \leq 1$.
By convention, we consider $B$ positive, hence $q>p$. 
Note that the model (\ref{action}) can be seen as a $(q,p)$-deformed matrix theory
with dual parameters  $q>1$ and $p<1$. The GWm is recovered as $q\to 1$ and $p\to 1$ ($B\to 0$).
Although this deformation  is not in the ordinary sense of Chakrabarti and co-workers \cite{cha,bur},
the model renormalizability suggests that
the deformed quantum algebras which are encountered in quantum group theory 
and quantum mechanics combined with nonlocal geometries may also be renormalizable. 
This is encouraging for the $q$-bosons studies 
and related theoretical models \cite{bur,BGH1,BGH2}.

The  bare propagator in the matrix base, at $\Omega=1$, is
\begin{equation} \label{propafixed}
C_{m n;k l} = C_{m n} \delta_{m l}\delta_{n k}, \qquad
C_{m n}= \frac{1}{A+q\,m+p\, n}\  ,
\end{equation}
and we use notations
\begin{equation}
\delta_{ml} = \delta_{m_1l_1} \delta_{m_2l_2}\ , \qquad q m+p n =q(m_1 + m_2) + p(n_1 + n_2) \ .
\end{equation}

Feynman rules involve only orientable graphs
with propagators oriented from $\bar{\phi}$ to $\phi$.
Arrows occur in alternating cyclic order at every vertex. 
For a field $\bar{\phi}_{m n}$, we call the index $m$ a 
{\it left index} and $n$ a {\it right index}. Consequently for the field $\phi_{kl}$, $k$ is a {\it right index} 
and $l$ a {\it left index}.

In \cite{Riv2}, the renormalizability of the model has been proved
in  the direct space at all order of perturbation. The renormalization of the four point function is essentially
the same as the one of the real GWm. But the two point function renormalization is more subtle due to the
left/right asymmetry of the model.

\section{Coupling constant flow}

We denote by $\Gamma^4(m,n,k,l)$ the amputated one particle irreducible (1PI) four point function 
with external indices $m,n,k,l$, and $\Sigma(m,n)$ the amputated 1PI two point function 
with external indices $m,n$ (the self-energy). To define the wave function renormalization,
we have to distinguish the left and right side of the ribbon and to attribute to each 
side its renormalization through the definitions
\begin{equation}
Z_L= 1-\frac{1}{q}\partial_L \Sigma(0,0),\qquad  Z_R= 1-\frac{1}{p}\partial_R \Sigma(0,0)
\end{equation}
which are  the derivative of the self-energy with respect to left and right indices. 
The wave function renormalization is then $Z=\sqrt{Z_L Z_R}$ corresponding
to a field rescaling $\phi \to Z^{1/2} \phi$. Therefore the effective coupling is
defined as 
\begin{equation}
\lambda^{eff}  = - \frac{\Gamma^{4} }{Z^2}  = - \frac{\Gamma^{4} }{Z_L Z_R}.
\end{equation}
\begin{theorem}
\label{theo}
The equation
\begin{eqnarray}\label{beautiful}
\Gamma^{4}(0,0,0,0)=- \lambda \, (1- \frac{1}{p}  \partial_R\Sigma(0,0))(  1 -\frac{1}{q} \partial_L\Sigma(0,0)),
\end{eqnarray}
where $\lambda$ is the bare constant, holds up to irrelevant terms  to  all orders of perturbation theory. 
\end{theorem}
Irrelevant terms have to be understood with respect to power counting and include in particular
all contributions of non-planar or planar graphs with more than one broken face.
This theorem is proved in the remaining of this section
following the method and ideas of \cite{diser2} adapted to the left
and right asymmetry.

\subsection{$(q,p)$-Ward identities}

The proof of Theorem \ref{theo} involve Ward identities (WI's)
related to  the $U(N)$ covariance of the theory.
These WI's can be extended to a class of 
classical or quantum symmetry transformations (translations and dilatations) letting the action invariant
up to a total derivative \cite{Gerh,BH}.
The following lemma holds.
\begin{lemma}
The planar one broken external face correlation functions satisfy
\begin{eqnarray}
&q (a-b)\langle[ \bar{\phi}\phi]_{a b} \phi_{\nu a} \bar{\phi}_{b \nu }\rangle_c=
\langle\phi_{\nu b} \bar{\phi}_{b \nu}\rangle_c
-\langle\bar{\phi}_{a \nu} \phi_{\nu a}\rangle_c, 
\label{ward2point}\\
&p (a-b) \langle[ \phi\bar{\phi}]_{a b} \phi_{b \mu} \bar{\phi}_{\mu a}\rangle_c= 
 \langle\bar{\phi}_{ \mu b} \phi_{b \mu}\rangle_c - \langle\phi_{a \mu} \bar{\phi}_{\mu a}\rangle_c,
\label{wi2}\\
&  q(a-b) \langle\phi_{\alpha a}[\bar{\phi}\phi]_{ab}\bar{\phi}_{b\nu}\phi_{\nu \delta}\bar{\phi}_{\delta \alpha}\rangle_c=
\langle\phi_{\alpha b}\bar{\phi}_{b \nu}\phi_{\nu \delta}\bar{\phi}_{\delta\alpha}\rangle_c-
\langle\phi_{\alpha a}\bar{\phi}_{a \nu}\phi_{\nu \delta}\bar{\phi}_{\delta\alpha}\rangle_c ,
\label{ward4point}\\
&
p (a-b) \langle\bar\phi_{\alpha a}[\phi\bar{\phi}]_{ab}\bar{\phi}_{b\nu}\phi_{\nu \delta}\bar{\phi}_{\delta \alpha}\rangle_c=
\langle\phi_{\alpha b}\bar{\phi}_{b \nu}\phi_{\nu \delta}\bar{\phi}_{\delta\alpha}\rangle_c 
-\langle\phi_{\alpha a}\bar{\phi}_{a \nu}\phi_{\nu \delta}\bar{\phi}_{\delta\alpha}\rangle_c.
\label{fwi}
\end{eqnarray}
\end{lemma}
The rest of this subsection is devoted to the proof of this Lemma.

\noindent{\bf  U(N) transformations.} 
Let $B$ be an infinitesimal hermitian matrix and consider the $U(N)$ group element $U=e^{\imath B}$
acting on the right and left on the matrix fields
\begin{eqnarray}
&&(\mbox{right})\quad \quad \phi^U:=\phi U;\;\;\;\; \bar{\phi}^U=U^{\dagger}\bar{\phi}\, ,
\label{right}\\
&&(\mbox{left})\quad \quad \phi^U:= U\phi ;\;\;\;\; \bar{\phi}^U=\bar{\phi}U^{\dagger} \, .\label{left}
\end{eqnarray}
The variation of the action under (\ref{right}) and (\ref{left})
is, at first order in $B$,
\begin{eqnarray}
\delta_L S=\imath B\big{(}X_L\bar{\phi}\phi-\bar{\phi}\phi X_L \big{)}, \quad
\delta_R S=\imath B\big{(} -X_R\phi\bar\phi+\phi\bar\phi X_R \big{)},
\end{eqnarray}
respectively. Similarly the variations of external sources are at first order
\begin{eqnarray}
\delta_L F=\imath B\big{(}-\bar{\phi}\eta+\bar{\eta}\phi{)} ,\quad \delta_R F=
\imath B\big{(}-\eta\bar{\phi}+\phi\bar{\eta} {)} .
\end{eqnarray}
As a consequence of the theory covariance, we have
\begin{eqnarray}
&&\frac{\delta_L \ln Z}{\delta B_{b a}}=0=
\frac{1}{Z(\bar{\eta},\eta)}\int d\bar{\phi} d\phi
   \big{(}-\frac{\delta_L S}{\delta B_{b a}}+\frac{\delta_L F}{\delta B_{b a}}\big{)}e^{-S+F}\nonumber\\
   &&=\frac{1}{Z(\bar{\eta},\eta)}\int d\bar{\phi} d\phi  ~e^{-S+F}
\big{(}-[X_L \bar{\phi}\phi-\bar{\phi}\phi X_L]_{a b}+
       [-\bar{\phi}\eta+\bar{\eta}\phi]_{a b}\big{)} ;\\
&&\frac{\delta_R \ln Z}{\delta B_{b a}}=0=
\frac{1}{Z(\bar{\eta},\eta)}\int d\bar{\phi} d\phi
   \big{(}-\frac{\delta_R S}{\delta B_{b a}}+\frac{\delta_R F}{\delta B_{b a}}\big{)}e^{-S+F}\nonumber\\
   &&=\frac{1}{Z(\bar{\eta},\eta)}\int d\bar{\phi} d\phi  ~e^{-S+F}
\big{(}-[-X_R {\phi}\bar\phi+{\phi}\bar\phi X_R]_{a b}+
       [-\eta\bar{\phi}+\phi\bar{\eta}]_{a b}\big{)}  .
\end{eqnarray}
\noindent {\bf Two point function Ward identities.} Applying the operator
$\partial_{\eta}\partial_{\bar{\eta}}|_{\eta=\bar{\eta}=0}$ 
on the above expressions and analyzing the result in terms of connected components leads to
\begin{eqnarray}
&0=\langle\partial_{\eta}\partial_{\bar{\eta}}\big{(}
 -[X_L \bar{\phi}\phi-\bar{\phi}\phi X_L]_{a b}+
       [-\bar{\phi}\eta+\bar{\eta}\phi]_{a b}\big{)}e^{F(\bar{\eta},\eta)} |_0\rangle_c \ ,
\label{derwi1}\\
&0=\langle\partial_{\eta}\partial_{\bar{\eta}}\big{(}
[X_R {\phi}\bar\phi-{\phi}\bar\phi X_R]_{a b}+
       [-\eta\bar{\phi}+\phi\bar{\eta}]_{a b}\big{)}e^{F(\bar{\eta},\eta)} |_0\rangle_c ,
\label{derwi2}
\end{eqnarray}
from which one deduces
\begin{eqnarray}
&\langle\frac{\partial(\bar{\eta}\phi)_{a b}}{\partial \bar{\eta}}\frac{\partial(\bar{\phi}\eta)}{\partial \eta}
-\frac{\partial(\bar{\phi}\eta)_{a b}}{\partial \eta}\frac{\partial (\bar{\eta}\phi)}{\partial \bar{\eta}}
- [X_L \bar{\phi}\phi-\bar{\phi}\phi X_L]_{a b}
\frac{\partial(\bar{\eta}\phi)}{\partial \bar{\eta}}\frac{\partial (\bar{\phi}\eta)}{\partial\eta}\rangle_c=0 ;\\
&\langle\frac{\partial(\phi\bar{\eta})_{a b}}{\partial \bar{\eta}}\frac{\partial(\bar{\phi}\eta)}{\partial \eta}
-\frac{\partial(\eta\bar{\phi})_{a b}}{\partial \eta}\frac{\partial (\bar{\eta}\phi)}{\partial \bar{\eta}}
+ [X_R \phi\bar{\phi}-\phi\bar{\phi} X_R]_{a b}
\frac{\partial(\bar{\eta}\phi)}{\partial \bar{\eta}}\frac{\partial (\bar{\phi}\eta)}{\partial\eta}\rangle_c=0 .
\end{eqnarray}
By the definition of $X_{L,R}$, we get
\begin{eqnarray}
&&q\, (a-b)\langle[ \bar{\phi}\phi]_{ab}
\frac{\partial(\bar{\eta}\phi)}{\partial \bar{\eta}}\frac{\partial (\bar{\phi}\eta)}{\partial\eta}\rangle_c=
\langle\frac{\partial(\bar{\eta}\phi)_{ab}}{\partial \bar{\eta}}\frac{\partial(\bar{\phi}\eta)}{\partial \eta}\rangle_c
-\langle\frac{\partial(\bar{\phi}\eta)_{ab}}{\partial \eta}\frac{\partial (\bar{\eta}\phi)}{\partial \bar{\eta}}\rangle, \nonumber\\
&&-p\,(a-b)\langle[\phi \bar{\phi}]_{ab}
\frac{\partial(\bar{\eta}\phi)}{\partial \bar{\eta}}\frac{\partial (\bar{\phi}\eta)}{\partial\eta}\rangle_c=
\langle\frac{\partial(\phi\bar{\eta})_{ab}}{\partial \bar{\eta}}\frac{\partial(\bar{\phi}\eta)}{\partial \eta}\rangle_c
-\langle\frac{\partial(\eta\bar{\phi})_{ab}}{\partial \eta}\frac{\partial (\bar{\eta}\phi)}{\partial \bar{\eta}}\rangle \ ,
\nonumber
\end{eqnarray}
and fixing $\bar{\eta}_{ \beta \alpha}$ and  $\eta_{ \nu \mu}$,  the previous relations become
\begin{eqnarray}
&& q\, (a-b)\langle[ \bar{\phi}\phi]_{a b} \phi_{\alpha \beta} 
\bar{\phi}_{\mu \nu }\rangle_c=
\langle\delta_{a \beta}\phi_{\alpha b} \bar{\phi}_{\mu \nu}\rangle_c
-\langle\delta _{b \mu }\bar{\phi}_{a \nu} \phi_{\alpha \beta}\rangle_c,
\label{fourpt1}\\
&&-p\,(a-b)\langle[\phi \bar{\phi}]_{a b} \phi_{\alpha \beta} 
\bar{\phi}_{\mu \nu }\rangle_c=
\langle\delta_{b \alpha}\phi_{a \beta } \bar{\phi}_{\mu \nu}\rangle_c
-\langle\delta _{a \nu }\bar{\phi}_{ \mu b} \phi_{\alpha \beta}\rangle_c.
\label{fourpt2}
\end{eqnarray}
Restricting to planar with a single 
external face terms requires 
$[\alpha=\nu,\;a=\beta,\; b=\mu]$ and $[\mu= \beta,\; \nu=a,\;b=\alpha ]$
for (\ref{fourpt1}) and (\ref{fourpt2}), respectively, and leads to (\ref{ward2point}) -(\ref{wi2}).

\noindent{\bf Four point function Ward identities.} 
Derivating further (\ref{derwi1}) and (\ref{derwi2}) yields
\begin{eqnarray}
&&q(a-b)\langle[\bar{\phi}\phi]_{a b}\partial_{\bar{\eta}_1}(\bar{\eta}\phi)
\partial_{\eta_1}(\bar{\phi}\eta) \partial_{\bar{\eta}_2}(\bar{\eta}\phi)
\partial_{\eta_2}(\bar{\phi}\eta) \rangle_c=  \\
&&\langle\partial_{\bar{\eta}_1}(\bar{\eta}\phi)
\partial_{\eta_1}(\bar{\phi}\eta)\big{[}
 \partial_{\bar{\eta_2}}
 (\bar{\eta}\phi)_{ab}\,\partial_{\eta_2}(\bar{\phi}\eta)-\partial_{\eta_2}(\bar{\phi}\eta)_{a b}\,
 \partial_{\bar{\eta}_2}(\bar{\eta}\phi) \big{]}\rangle_c+1 \leftrightarrow 2 ;
\nonumber\\
&&-p(a-b)\langle[\phi\bar{\phi}]_{a b}\,\partial_{\bar{\eta}_1}(\bar{\eta}\phi)
\partial_{\eta_1}(\bar{\phi}\eta) \partial_{\bar{\eta}_2}(\bar{\eta}\phi)
\partial_{\eta_2}(\bar{\phi}\eta) \rangle_c= \\
&&\langle\partial_{\bar{\eta}_1}(\bar{\eta}\phi)
\partial_{\eta_1}(\bar{\phi}\eta)\big{[}
 \partial_{\bar{\eta_2}}
 (\phi\bar{\eta})_{ab}\partial_{\eta_2}(\bar{\phi}\eta)-\partial_{\eta_2}(\eta\bar{\phi})_{a b}\,
 \partial_{\bar{\eta}_2}(\bar{\eta}\phi) \big{]}\rangle_c+1 \leftrightarrow 2 \ .\nonumber
\end{eqnarray}
A straightforward derivation at fixed $\bar{\eta}_{1,\beta \alpha}$, $\eta_{1, \nu\mu}$, 
$\bar{\eta}_{2,\delta \gamma}$ and $\eta_{2,\sigma \rho}$ gives
\begin{eqnarray}
&&q(a-b)\langle[\bar{\phi}\phi]_{ab}\,\phi_{\alpha \beta}\bar{\phi}_{\mu \nu}\phi_{\gamma \delta}
\bar{\phi}_{\rho \sigma}\rangle_c =\langle\phi_{\alpha \beta}\bar{\phi}_{\mu \nu}  \delta_{a \delta}\phi_{\gamma b}\bar{\phi}_{\rho \sigma}\rangle_c
\\
&& -\langle\phi_{\alpha \beta}\bar{\phi}_{\mu \nu}\phi_{\gamma \delta}\bar{\phi}_{a \sigma}\delta_{b \rho}\rangle_c +
\langle\phi_{\gamma \delta}\bar{\phi}_{\rho \sigma}  \delta_{a \beta}\phi_{\alpha b}\bar{\phi}_{\mu \nu}\rangle_c
-\langle\phi_{\gamma \delta}\bar{\phi}_{\rho \sigma}\phi_{\alpha \beta}\bar{\phi}_{a \nu}\delta_{b \mu}\rangle_c \ ;
\nonumber\\
&&-p(a-b)\langle[\phi\bar\phi]_{ab}\,\phi_{\alpha \beta}\bar{\phi}_{\mu \nu}\phi_{\gamma \delta}
\bar{\phi}_{\rho \sigma}\rangle_c=\langle\phi_{\alpha \beta}\bar{\phi}_{\mu \nu}  \delta_{\gamma b}
\phi_{a \delta}\bar{\phi}_{\rho \sigma}\rangle_c
\\&& - \langle\phi_{\alpha \beta}\bar{\phi}_{\mu \nu}\phi_{\gamma \delta}\bar{\phi}_{ \rho b}\delta_{a \sigma}\rangle_c
+ \langle\phi_{\gamma \delta}\bar{\phi}_{\rho \sigma}  \delta_{\alpha b}\phi_{a \beta}\bar{\phi}_{\mu \nu}\rangle_c
-\langle\phi_{\gamma \delta}\bar{\phi}_{\rho \sigma}\phi_{\alpha \beta}\bar{\phi}_{ \mu b}\delta_{a \nu}\rangle_c \ .
\nonumber
\end{eqnarray}
Neglecting irrelevant graphs gives (\ref{ward4point})-(\ref{fwi}), completing 
the proof of the lemma.
\hfill$\square$

\begin{figure}[hbt]
\centerline{
\includegraphics[width=10cm]{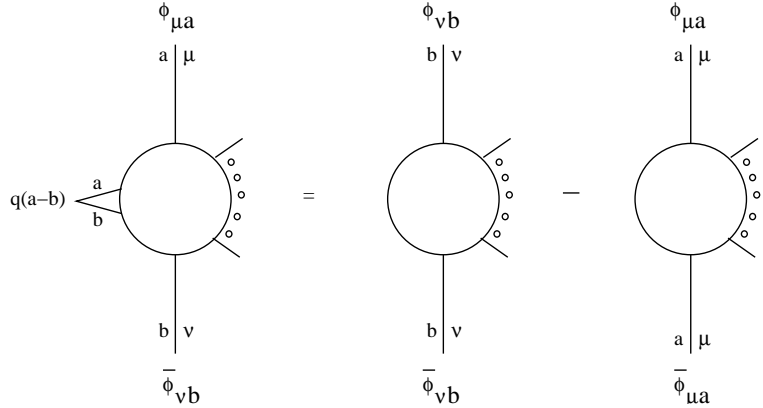}
}
\caption{$q$-Ward identity for a $2\ell$ point function with insertion on the left face.}\label{fig:Ward}
\end{figure}

A simple induction proves that such identities hold for 
$2\ell$ point functions with a left or right insertion, and for any integer $\ell$,
as depicted in Fig.\ref{fig:Ward}.

\subsection{Proof of the theorem}

\begin{figure}[hbt]
\centerline{
\includegraphics[width=120mm]{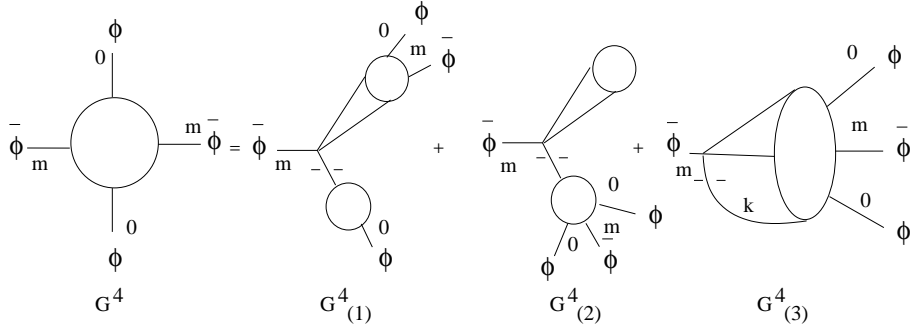}
}
\caption{The left Dyson equation.}\label{fig:dyson}
\end{figure}

Besides the WI's, the proof of 
Theorem \ref{theo}  uses  left (right) DSe  for
four point functions  with the two left (right) indices
equal to $m$
(see Fig.\ref{fig:dyson} for the left DSe), namely
\begin{eqnarray}
&&G^{4}(0,m,0,m)= G_{(1)}^{4}(0,m,0,m) + G_{(2)}^{4}(0,m,0,m) + G_{(3)}^{4}(0,m,0,m),
\label{Dyson}\\
&&G^{4}(m,0,m,0)= G_{(1)}^{4}(m,0,m,0) + G_{(2)}^{4}(0,m,0,m) + G_{(3)}^{4}(m,0,m,0),
\label{rdys}
\end{eqnarray}
where $G^{4}(m,n,k,l)$ is the connected planar single external face
four point function. Equation (\ref{Dyson}) is
the left DSe whereas (\ref{rdys}) is the right DSe.  A clear comment of the meaning
of these relations is provided in \cite{diser2}.
The term  $G^{4}_{(2)}$ is zero by mass renormalization.

Denote $\partial_L F(m,n):= \partial_{m_i} F(m,n)$
and $\partial_R F(m,n):= \partial_{n_i} F(m,n)$, where $m$ and $n$ are respectively left and
right indices. The following lemma holds.
\begin{lemma}\label{lem}
Up to irrelevant terms we have
\begin{eqnarray}
&&G_{(1)}^{4}(0,m,0,m)= -\lambda (G^{2}(0,m))^4\left(Z_R +\frac{A_r\; \partial_R \Sigma(0,0) }{p\,(pm + A_r)}\right)
Z_L,\label{gr41}\\
&&G_{(1)}^{4}(m,0,m,0)= -\lambda (G^{2}(m,0))^4\left(Z_L +\frac{A_r \;\partial_L \Sigma(0,0) }{q\,(qm + A_r)}   \right) Z_R,\\
&&G_{(3)}^{4}(0,m,0,m)= -G^4(0,m,0,m) \frac{A_r}{(pm+A_r)} \frac{\partial_R \Sigma^{R}(0,0)}{(p-\partial_R \Sigma(0,0)},
\label{gr43}\\
&&G_{(3)}^{4}(m,0,m,0)=  -G^4(m,0,m,0) \frac{A_r}{(qm+A_r)} \frac{\partial_L \Sigma^{L}(0,0)}{(q-\partial_L \Sigma(0,0)}.
\end{eqnarray}
where $\Sigma^{R,L}(0,0)$ is defined in (\ref{lostct}) below
and $G^2(m,n)$ is the connected planar one broken face two point function.
\end{lemma}
{\bf Proof.} Let us prove (\ref{gr41}) and (\ref{gr43}).
The proof of the other expressions is analogous.

$G^2(m,n)$ is given by a well known sum of a geometric series 
\begin{eqnarray}
\label{G2Sigmarelation}
 G^{2}(m,n)=\frac{C_{m n}}{1-C_{m n}\Sigma(m,n)}=\frac{1}{C_{m n}^{-1}-\Sigma(m,n)} \, .
\end{eqnarray} 
Let $G^L _{ins}(a,b;...)$ be the planar one broken face 
connected function with one index jump on the left
from $a$ to $b$. 
Using (\ref{ward2point}), one writes
\begin{eqnarray}
q\,(a-b) ~ G^{2,L}_{ins}(a,b;\nu)=G^{2}(b,\nu)-G^{2}(a,\nu)\, .
\label{wif}
\end{eqnarray} 

We note that WI's and the DSe  have a  meaning both
in the bare (of mass $A=A_{bare}$) and in the mass renormalized theory ($A_{r}= A_{bare} - \Sigma(0,0)$).
The latter case implies that  every two point 1PI subgraph should be subtracted at 0 external indices.
In the following, we use the mass-renormalized derivation\footnote{An equivalence with
the bare theory could be deduced from \cite{diser2}.}. The mass renormalized theory 
is free from quadratic divergences. Residual logarithmic divergences in the UV cutoff can be 
read off the effective series as argued in \cite{diser,lRiv}. 

$G^{4}_{(1)}$ decomposes as
\begin{eqnarray}
G^4_{(1)}(0,m,0,m)=-\lambda C_{0 m} G^{2}(0, m) G^{2,L}_{ins}(0,0;m)\,.
 \end{eqnarray}

By the WI (\ref{wif}), we obtain
\begin{eqnarray}
G^{2,L}_{ins}(0,0;m)&=&\lim_{\zeta\rightarrow 0}G^{2}_{ins}(\zeta ,0;m)=
\frac{1}{q}\lim_{\zeta\rightarrow 0}\frac{G^{2}(0,m)-G^{2}(\zeta,m)}{\zeta}\nonumber\\
&=&-\frac{1}{q}\,\partial_{L}G^{2}(0,m) \, .
\end{eqnarray}

Using the  form (\ref{G2Sigmarelation}) of $G^{2}(0,m)$,  we get
\begin{eqnarray}\label{g41}
G^4_{(1)}(0,m,0,m)&=& - \frac{\lambda}{q}\,
C_{0m}\frac{C_{0m}C^2_{0m}\left[q-\partial_{L}\Sigma(0,m)\right]}{\left[1-C_{0m}\Sigma(0,m)\right]
(1-C_{0m}\Sigma(0,m))^2}\nonumber\\
&=&-\frac{\lambda}{q}\left[G^{2}(0,m)\right]^{4}\frac{C_{0m}}{G^{2}(0,m)}\left[q-\partial_{L}\Sigma(0,m)\right]\, .
\end{eqnarray}

A Taylor expansion gives the self energy up to irrelevant terms \cite{GW2},
\begin{eqnarray}
\label{PropDressed}
\Sigma(m,n)=\Sigma(0,0)+m\,\partial_L\Sigma(0,0)+n\,\partial_R\Sigma(0,0) .
\end{eqnarray} 
Keeping in mind that  $C^{-1}_{0m}=pm+A_{r}$, we have (again  up to irrelevant terms)
\begin{eqnarray}
\label{G2(0,m)}
G^{2}(0,m)=\frac{1}{pm+A_{bare}-\Sigma(0,m)}=\frac{1}{m\left[p-\partial_R\Sigma(0,0)\right]+A_{r}}
\, ,
\end{eqnarray}
and
\begin{eqnarray} \label{cdressed}
\frac{C_{0m}}{G^{2}(0,m)}=\frac{1}{p}(p-\partial_R\Sigma(0,0))+\frac{A_{r}}{p(pm+A_{r})}\partial_R\Sigma(0,0) \, .
\end{eqnarray}

Substituting  (\ref{cdressed}) in (\ref{g41}) we get
\begin{eqnarray}
\label{g41final}
G^4_{(1)}(0,m,0,m)&=&-\lambda \left[G^{2}(0,m)\right]^{4}\left( 
\frac{1}{p}(p-\partial_R\Sigma(0,0))+\frac{A_{r}}{p(pm+A_{r})}\partial_R\Sigma(0,0)
\right) \nonumber\\
&&\left[\frac{1}{q}(q-\partial_{L}\Sigma(0,m))\right]\, .
\end{eqnarray}

\begin{figure}[hbt]
\centerline{
\includegraphics[width=140mm]{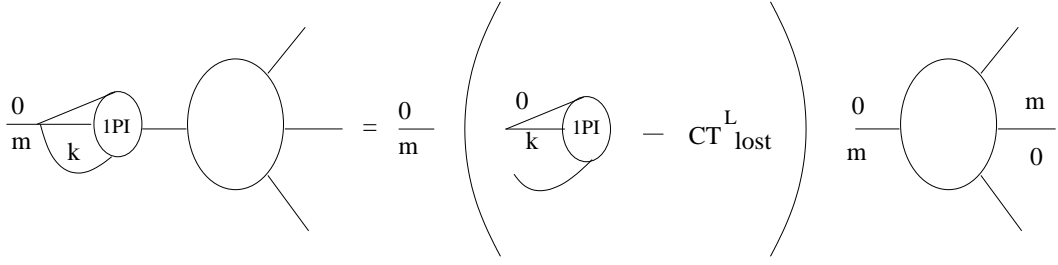}
}
\caption{Two point left insertion and opening of the loop with index $k$.}\label{fig:insertion}
\end{figure}
\begin{figure}[hbt]
\centerline{
\includegraphics[width=90mm]{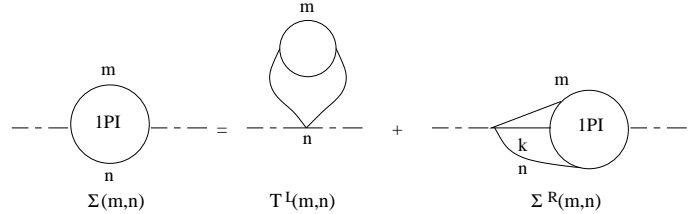}
}
\caption{The self energy.}\label{fig:selfenergy}
\end{figure}

To evaluate $G^4_{(3)}(0,m,0,m)$, we need to ``open" the face ``on the right'' 
in the $k$ loop in the third term of Fig.\ref{fig:dyson}. 
The left bare correlation functions are  given by
\begin{eqnarray}
\label{opening}
G^{4,bare}_{(3)}(0,m,0,m)=-\lambda C_{0m}\sum_{ k} G^{4,bare,L}_{ins}(k,0;m,0,m)\, .
\end{eqnarray}
The face indexed by $k$ may
belong to a  1PI two point insertion in $G^{4}_{(3)}$ (see Fig.\ref{fig:insertion}). 
In that case, because we use the mass renormalized expansion,  one has to introduce a counterterm
in order to compensate the one lost during the ``opening" process. In other terms, we have
\begin{equation}
\label{Open2}
G^4_{(3)}(0,m,0,m)= - \lambda C_{0m}\sum_{k} G^{4,L}_{ins}(0,k;m,0,m) -  C_{0m}(CT^L_{lost})G^{4}(0,m,0,m)\,.
\end{equation}
It turns out that all two point function counterterms contribute to $CT^L_{lost}$ except those
of the generalized left  tadpole. We write
\begin{eqnarray}
\label{eq:leftright}
\Sigma(m,n)=T^{L}(m,n)+\Sigma^R(m,n)\,
\end{eqnarray}
with $T^{L}$ the  generalized left tadpole contribution and
$\Sigma^R$ the rest.
$T^{L}(m,n)$ is a left border insertion hence
does not depend upon the right index $n$ (see Fig.\ref{fig:selfenergy}).

With these notations the missing mass counterterm is given by
\begin{eqnarray}\label{lostct}
CT^L_{lost}=\Sigma^R(0,0)=\Sigma(0,0)-T^{L}\, .
\end{eqnarray}
To compute $\Sigma^{R}(0,0)$, we open its face indexed by $k$ 
and use (\ref{ward2point})  to get
\begin{eqnarray}
\Sigma^R(0,0)&=&- \frac{\lambda}{G^{2}(0,0)}\sum_{k}G^{2,L}_{ins}(0,k;0)
    =- \frac{\lambda}{q}\frac{1}{G^{2}(0,0)}\sum_{k}\frac{1}{k}[G^2(0,0)-G^2(k,0)]\cr
   & =&-\frac{\lambda}{q}\sum_{k}\frac{1}{k} \biggl(1 -\frac{G^{2}(k,0)}{G^{2}(0,0)}\biggr) \, .
\label{S2} 
\end{eqnarray}
Then (\ref{Open2}) and (\ref{S2}) imply that
\begin{eqnarray}\label{S3}
G^4_{(3)}(0,m,0,m)&=& - \lambda C_{0m}\sum_{k} G^{4,L}_{ins}(0,k;m,0,m)\nonumber\\
&-&\frac{(- \lambda)}{q} \,C_{0m} G^{4}(0,m,0,m) \sum_{k}\frac{1}{k}  
\biggl( 1- \frac{G^{2}(k,0)}{G^2(0,0)}  \biggr)  \, .
\end{eqnarray}
Equation (\ref{ward4point}) reexpresses the first term in (\ref{S3})
\begin{eqnarray}
\label{Ward4}
 - \lambda C_{0m} \sum_{k} G^{4,L}_{ins}(0,k;m,0,m)= - 
\frac{\lambda}{q}\, C_{0m} \sum_{k} \frac{1}{k}\biggl( G^{4}(0,m,0,m)-G^{4}(k,m,0,m) \biggr)\, .
\end{eqnarray}
The second term in (\ref{Ward4}) is at least cubic in $k$, hence irrelevant. 
The above sums over $k$ for  $G^4(k,m,0,m)$ are always convergent (see \cite{diser2}).
We inject (\ref{Ward4}) in (\ref{S3}) and obtain
\begin{eqnarray}
\label{G3}
G^4_{(3)}(0,m,0,m) =-\frac{\lambda}{q} \, C_{0m}\frac{G^{4}(0,m,0,m)}{G^2(0,0)}\sum_{k}
\frac{G^{2}(k,0)}{k} \, .
\end{eqnarray}
From (\ref {G2(0,m)}), we obtain
\begin{eqnarray}
\label{derivee}
\sum_{k}\frac{G^{2}(k,0)}{k}=\sum_{k}\frac{G^{2}(k,0)}{k}\bigl( \frac{1}{G^{2}(0,1)}-\frac{1}{G^{2}(0,0)}\bigr)
\frac{1}{(p-\partial_R\Sigma(0,0))}.
\end{eqnarray}
Performing the same manipulations as in (\ref{S2}), we express
\begin{eqnarray}
\label{S2new}
\Sigma^R(0,1)&=&- \frac{\lambda}{q}\,\sum_{k}\frac{1}{k} 
\biggl(1 -\frac{G^{2}(k,1)}{G^2(0,1)}\biggr) =- \frac{\lambda}{q}\,\sum_{k}\frac{1}{k} 
\biggl(1 -\frac{G^{2}(k,0)}{G^2(0,1)}\biggr)\, .
\end{eqnarray}
up to an irrelevant term. 
Substituting (\ref{S2}) and (\ref{S2new}) in (\ref{derivee}),
\begin{eqnarray}
- \lambda \sum_{k}\frac{G^{2}(k,0)}{k}=\frac{q\,(\Sigma^R(0,0)-\Sigma^R(0,1))}{p-\partial_R\Sigma(0,0)}
=-\frac{q\,\partial_{R}\Sigma^R(0,0)}{p-\partial_R\Sigma(0,0)}.
\end{eqnarray}
Therefore,
\begin{eqnarray}\label{g43}
G^4_{(3)}(0,m,0,m)&=&-C_{0m}G^{4}(0,m,0,m)\frac{1}{G^{2}(0,0)}\frac{\partial_R\Sigma^R(0,0)}{(p-\partial_R\Sigma(0,0))}
\nonumber\\
&=&-G^{4}(0,m,0,m)
\frac{A_{r} \; \partial_R\Sigma^R(0,0)}{(pm+A_{r}) (p-\partial_R\Sigma(0,0))} \ 
\end{eqnarray}
which achieves the proof of Lemma \ref{lem}.
\hfill$\square$ 

{\bf Proof of Theorem \ref{theo}.} 
Plugging (\ref{g41final}) and (\ref{g43}) in (\ref{Dyson}), one has
\begin{eqnarray}
\label{final}
&&G^4(0,m,0,m)\Big{(}1+
\frac{A_{r}\; \partial_R\Sigma^R(0,0)}{(pm+A_{r}) \; ( p-\partial_R\Sigma(0,0)) }\Big{)}
\\
&&=- \lambda (G^{2}(0,m))^{4}\Big{(} \frac{1}{p}(p-\partial_R\Sigma(0,0))+\frac{A_{r}}{p(pm+A_{r})}\partial_R\Sigma(0,0)\Big{)}
\frac{1}{q}(q-\partial_{L}\Sigma(0,m))\, .\nonumber
\end{eqnarray}
Multiplying (\ref{final}) by $(p-\partial_R\Sigma(0,0))/p$, amputating four times
and neglecting the irrelevant differences $\Gamma^4(0,m,0,m)-\Gamma^4(0,0,0,0)$ and 
$\partial_L\Sigma(0,m)-\partial_L\Sigma(0,0)$,  we finally find 
\begin{eqnarray}
\Gamma^{4}(0,0,0,0)=- \lambda(1-\frac{1}{q}\partial_L\Sigma(0,0)) (1-\frac{1}{p}\partial_R\Sigma(0,0))\, .
\end{eqnarray}
which completes the proof of (\ref{beautiful}).
\hfill$\square$ 

\section{One loop RG flow of $(q,p)$ parameters}

The left and right wave function renormalizations $Z_{L}$ and $Z_{R}$, respectively,
determine the RG flow of $q$ and $p$.  To compute RG flows,
we need to introduce some slice decomposition \cite{Riv}.
After renormalization of the
field $\phi \to \phi(Z_RZ_L)^{1/4}$, the discrete RG flow equation are
\begin{equation}
\lambda_{i-1}=\lambda_i, \qquad
q_{i-1} = q_i \left(\frac{Z_L}{Z_R}\right)^{\frac{1}{2}}, \qquad
p_{i-1} = p_i \left(\frac{Z_R}{Z_L}\right)^{\frac{1}{2}}.
\end{equation}
At one loop, only the planar "up" and "down"  tadpoles \cite{diser} contribute
to the self-energy $\Sigma(m,n)$. We get a factor of symmetry of $2$ so that 
\begin{equation}
\Sigma(m,n) = -\lambda \sum_{r \in \mathbb{N}^2} \left(C_{mr}+C_{rn}\right),\;\;
\end{equation}
where $C_{mr}$ and $C_{rn}$ are the bare propagators.

\begin{figure}
\hspace{0.5cm}
\put(0,75){$p$}\put(0,30){$q$}
\centering
\includegraphics[width=7cm]{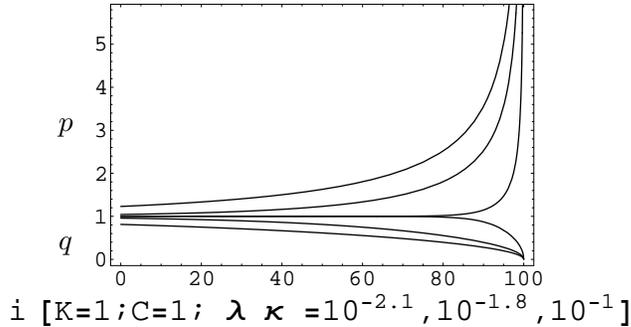}
\caption{{\small  RG flow of $q(i)$ and $p(i)$ versus $i$ with cutoff $\Lambda=100$
and $p_{uv}=10^{-6}$.}}
\label{repres2}
\end{figure}

A direct calculation yields, with $r \in \mathbb{N}^2$,
\begin{equation}
Z_{L} = 1- \lambda \sum_{r}  \frac{1}{( p r+ A)^2},  \quad
Z_{R} = 1- \lambda \sum_{r} \frac{1}{ (q r+ A)^2}.
\end{equation}
Hence, at first order in $\lambda$,
\begin{eqnarray}
\sqrt{\frac{Z_L}{Z_R}} = 1- \frac{1}{2}\lambda   \sum_{r} 
\biggr[ \frac{1}{( p r+ A)^2}-  \frac{1}{( q r+ A)^2} \biggl] + O(\lambda^2),
\end{eqnarray}
The logarithmically divergent part of these sums governs the flows. In a slice
corresponding to $r_1, r_2 \in [M^{i-1}, M^i]$, we have
\begin{equation}
 \sum_{r_1,r_2=M^{i-1}}^{M^{i}} \frac{1}{ (q r+ A)^2}
= \frac{1}{q^2} \sum_{r_1,r_2=M^{i-1}}^{M^{i}} \frac{1}{(r+
 \frac{A}{q})^2} = \frac{1}{q^2} \, \kappa + O (M^{-i})
\end{equation}
where the constant $\kappa$ is independent of $i$.
We obtain at one loop,
\begin{eqnarray}
q_{i-1} = q_i \left[ 1- \frac{\lambda_i}{2} \left(\frac{1}{p^2_i} - \frac{1}{q^2_i}\right) \kappa  \right] , \quad
p_{i-1} = p_i \left[ 1- \frac{\lambda_i}{2} \left(\frac{1}{q^2_i} - \frac{1}{p^2_i}\right)\kappa   \right]
\end{eqnarray}
from which the discrete flows are deduced
\begin{eqnarray}
\frac{d \,q_i}{d\, i} = \frac{\lambda_i}{2}  \,q_i \,\left( \frac{1}{p^2_i} - \frac{1}{q^2_i}   \right) \kappa,\quad
\frac{d \,p_i}{d\, i} =\frac{\lambda_i}{2} \,p_i \, \left( \frac{1}{q^2_i} - \frac{1}{p^2_i}   \right)\kappa.
\label{pqrg}
\end{eqnarray}
This leads directly to
\begin{eqnarray}
\frac{d \,q_i}{q_i \,d\, i} +  \frac{ d \,p_i}{p_i\, d\, i} =0 \quad\Leftrightarrow\quad
q_i\,p_i =K,
\end{eqnarray}
where $K$ is some positive constant.
We substitute $q_i=K/p_i$ in (\ref{pqrg}) and find the solutions
\begin{eqnarray}
&&p(i)^{2} =K  \frac{  e^{2 \lambda_i\,\kappa\,(\Lambda - i)}(p_{uv}^2/K+1)+p_{uv}^2/K-1}
{   e^{2 \lambda_i\,\kappa\,(\Lambda - i)}(p_{uv}^2/K+1)-(p_{uv}^2/K-1)} ,
\label{pqsol}\\
&&p_{uv}^{2} =-K  \frac{  e^{2 \lambda_i\,\kappa\,(\Lambda - i)}(p(i)^2/K-1)+p(i)^2/K+1}
{ e^{2 \lambda_i\,\kappa\,(\Lambda - i)}(p(i)^2/K-1)-(p(i)^2/K+1)} ,
\label{pqsol2}
\end{eqnarray}
where $\Lambda$ stands for the UV cutoff, and $p_{uv}$  the bare value of $p$
(see Appendix). Similar
expressions of $q_i$ and $q_{uv}$ follow.
Graphic representations of $p(i)$ and $q(i)$ versus $i$ for various values of the parameters
are given in Fig.\ref{repres2}. 

\section{Conclusion}
We have proved that the $\beta$-function governing the RG flow
of the coupling constant of the complex GWm with magnetic field
vanishes at all orders of perturbation.  We have also computed at one loop the RG flows of the new  wave function 
parameters $(q,p)$.  The non-Gaussian fixed point  $p=q$ lies on the IR side rather than the UV one.

The motivation for studying these models in magnetic field comes
from the quantum Hall effect physics, although this physics requires a different propagator and $2+1$ dimensions.
We hope to describe the Hall plateaux as fixed points of a noncommutative RG flow. The results for the particular 
toy model considered here may not seem encouraging at first sight. Indeed 
the only infinite direction of such noncommutative RG is the UV one and that's where 
we do not find fixed points.  However recall that the noncommutative interpretation of long and short distances
is subtle, IR and UV in NC really referring to low versus high energy.
The physics at small energies in a Hall fluid is well described in terms of anyons, whereas electrons appear
as high energy particles \cite{Gir}. Hence anyonic physics may be described by IR rather than UV fixed points.

\section*{Acknowledgments}

J. B. G. thanks  the ANR Program ``GENOPHY" and the  Daniel Iagolnitzer Foundation (France)
for a research grant in the LPT-Orsay, Paris Sud XI.  

\section*{Appendix}
 \renewcommand{\theequation}{A.\arabic{equation}}
\setcounter{equation}{0}

Consider the linear differential equation given by (\ref{pqrg})
after substituting $q_i=K/p_i$, $p_i=p(i)$ and $q_i=q(i)$, 
\begin{eqnarray}
\frac{d \,p(i)}{d\, i} =\frac{\lambda_i}{2} \,p(i) \, \left( \frac{p(i)^2}{K^2} - \frac{1}{p^2(i)}   \right)\kappa. 
\end{eqnarray}
Separating variables and putting $u(i)=p(i)^2$, we get
\begin{eqnarray}
K\,\frac{d\,u}{u^2-1} =  \lambda_i\,\kappa \,\,d\,i 
\end{eqnarray}
which can be easily integrated. Let us remark
that $\lambda_i \simeq \lambda$, Theorem \ref{theo}
having proved that the flow of $\lambda_i$ is actually bounded.
Between the $i^{th}$ slice and the UV cutoff $\Lambda$, $p(i)$ varies
from $p(i)$ to its bare value $p_{uv}$, so that
\begin{eqnarray}
\frac{1}{2}\left\{ \ln\left(\frac{p_{uv}^2/K-1}{p_{uv}^2/K+1} \right) - \ln\left(\frac{p(i)^2/K-1}{p(i)^2/K+1} \right) \right\}
= \lambda\,\kappa \,\ (\Lambda - i).
\end{eqnarray}
The bare and running values of $p$ are related through
\begin{eqnarray}
\frac{p(i)^2/K+1}{p(i)^2/K-1}= e^{2\lambda \kappa(\Lambda-i)} \; \frac{p_{uv}^2/K+1}{p_{uv}^2/K-1},
\end{eqnarray}
and we recover (\ref{pqsol}) and (\ref{pqsol2}).

\end{document}